\documentstyle[12pt,aasms4]{article}

\begin{document}

\title{Average Cosmological Invariant Parameters of Cosmic Gamma Ray Bursts}
\author{Igor G.~Mitrofanov$^1$, Dmitrij S.~Anfimov$^1$, Maxim L.~Litvak$^1$ \and %
Michael S.~Briggs$^2$, William S.~Paciesas$^2$, Geoffrey N.~Pendleton$^2$,
Robert D.~Preece$^2$ \and Charles A.~Meegan$^3$}
\affil{$^1$Space Research Institute, Profsojuznaya str. 84/32, 117810
Moscow, Russia} 
\affil{$^2$Department of Physics, University of Alabama in Huntsville,
Huntsville, AL 35899} 
\affil{$^3$NASA/Marshall Space Flight Center, Huntsville, AL 35812}

\begin{abstract}
Average cosmological invariant parameters (ACIPs) are calculated for six
groups of BATSE cosmic gamma-ray bursts selected by their peak fluxes on the
1.024~s time scale. The ACIPs represent the average temporal and spectral
properties of these events equally in the observer frame of reference and in
the co-moving frames of outbursting emitters. The parameters are determined
separately for rise fronts and for back slopes of bursts, defined as the
time profiles before and after the main peaks, respectively. The ACIPs for
the rise fronts are found to be different for different intensity groups,
while the ACIPs for the back slopes show no significant dependence on
intensity. We conclude that emitters of bursts manifest standard average
properties only during the back slopes of bursts.
\end{abstract}

\section{Introduction}

The isotropy of gamma-ray bursts on the sky combined with the significant
deviation of the log$N$/log$P$ curve from a $-3/2$ power law (\cite{BATSE})
gave the first clear evidence that sources of gamma-ray bursts are at
cosmological distances, where the deficit of observed number of dim events
is a direct consequence of the non-Euclidean-nuture of the expanding
Universe (\cite{Pasz1}). Strong evidence in favor of the cosmological
paradigm has recently been provided by the detection of red-shifted spectral
lines from the optical counterparts of GRB 970508 (\cite{Metzger97}) and GRB
971214 (\cite{Kulkarni98}).

As soon as the cosmological paradigm became popular for gamma-ray bursts,
two effects were suggested to test it using the observational data of
gamma-ray bursts. The first one is the effect of cosmological red shift,
which predicts that the energy of spectral features of dimmer bursts should
be red-shifted in respect to similar features of bright events. The second
one is the complementary effect of cosmological time dilation, which
predicts that light curves of dimmer bursts should be time-stretched with
respect to those of brighter events.

Since both effects are associated with the geometry of the expending
Universe, we will refer to them as ${\it geometrical}$ effects. The
transformation factors are known to be 
\begin{equation}
Y(z_{{\rm br}}, z_{{\rm dim}}) = (1 + z_{{\rm dim}}) / (1 + z_{{\rm br}}), 
\end{equation}
where $z_{{\rm br}}$ and $z_{{\rm dim}}$ are the red shifts of emitters of
bright and dim bursts, respectively.

If the brightest bursts are associated with the nearest objects and the
dimmest bursts are related to the furthest emitters, their intensities could
be used to determine distances to the corresponding emitters. In this case
the cosmological effects should lead to a hardness/intensity correlation
(due to the red shift), and a stretching/intensity anti-correlation (due to
the time dilation).

However, gamma-ray bursts are known to be very different from each other. In
searching for the generic effects of hardness/intensity correlations and
stretching/intensity anti-correlation between different intensity groups,
one has to identify generic signatures among the divergence of properties of
individual events to represent the properties of {\it typical} burst
emission. 

Using such signatures, the effects of hardness/intensity correlation and
stretching/intensity anti-correlation have been studied by comparing groups
of BATSE bursts with different brightnesses. The average peak energy $E_{%
{\rm p}}$ of $\nu F_\nu $ spectra was found to show a hardness/intensity
correlation consistent with cosmological red shift (\cite{Mall}), and there
is evidence that average emissivity curves of dimmer bursts are stretched
relative to those of brighter bursts, although the results of different
analyses are not entirely consistent (\cite{Norr1,Norr2,Mit4,Mit5}). The
possibility that the stretching of dimmer bursts might result from a
selection effect has been considered, and it has been shown that it is not a
selection effect  (\cite{Wij1}). On the other hand, it is also not
clear that the results of red shift and time dilation studies agree with
each other (\cite{Mit6}).

On the other hand, the logically simple concept that burst intensities are
``standard candles'' may not be correct. It has been shown (\cite{Brain})
that the broad spread of observed intensities of bursts could result from a
broad spread of intrinsic luminosities of emitters. Also, three bursts with
measured red-shifts of afterglowing optical emission have gamma-ray
luminosity differing over a range of $\sim$20 times. Despite of small sample
size, these three events provide quite good evidence that GRBs are not
standard candles. The correlations of burst hardness and duration with
intensity could be related in this case to the intrinsic properties of
sources rather than to geometrical effects of cosmological space (cite{Mit2}%
). Furthermore, one should take into account that the astronomical
population of emitters could be quite different at close and distant
cosmological distances. In this case the difference between brightness
groups could reflect the intrinsic evolution of sources rather than
cosmological effects.

Therefore, to test the cosmological paradigm it is necessary to separate in
the data the {\it geometrical effects} due to cosmological expansion of the
Universe from the {\it physical effects} due to intrinsic luminosity-based
correlation of emitters and the {\it astronomical effects} due to evolution
of sources over the different red shifts.

\section{Double-peak energy, as a burst spectral parameter}

To compare the average temporal and spectral signatures of gamma-ray bursts
with different brightness, 6 intensity groups, each consisting of $\sim 100$
bursts, were selected by their peak fluxes $F_{{\rm max}}$ on the 1024~ms
time scale in the 3B BATSE Catalog (\cite{3B}), as shown in Table 1.

The time profile of each burst has a well-defined moment $t_{{\rm max}}$
when the observed flux reaches a maximum $F_{{\rm max}}$. One might
postulate that peaks are associated with some particular physical
transition, when the average rising trend of intensity before the peak is
converted into the decaying tail after that. The procedure of peak alignment
has the physical sense to combine all bursts together at the same stage of
the emission process. On the other hand, the $\nu F_\nu $ energy spectrum of
burst emission usually has a well-defined peak energy $E_{{\rm p}}$, but $E_{%
{\rm p}}$ typically varies with time during a burst. One could, however,
introduce a single-valued spectral parameter for each burst, which is the
peak energy $E_{{\rm p,max}}$ of the spectrum at the time of peak photon
flux $F_{{\rm max}}$. We will refer to $E_{{\rm p,max}}$ as the {\it %
double-peak energy}, which corresponds to photons with the largest spectral
density of emission.

We evaluated the double-peak energy $E_{{\rm p,max}}$ for each BATSE burst
using the CONT data, which has a time resolution of 2048~ms (\cite{Anf}).
The distributions of these energies for the reference 
group and for the dimmest intensity group are shown in Figure 1. The observed distributions
for these groups might be compared with K-S test provided that the
reference group is shifted leftward to match the curves for the dimmer
groups. 
The lowest K-S probability is equal to 0.3 for the comparison between reference group and dimmest intensity group
when they are shifted to the same mean.
The present data allows us to measure with statistical significance
these shifts, but does not reveal any difference in the shapes. We might
assume that there is an {\it universal} log-normal law for all these
distributions, which can be parameterized using a set of red-shifting
factors $Y^{(i)}$, defined by shifting the log-normal distribution for the
brightest reference group ($i=1$) to provide the best fit for the
corresponding distributions of the dimmer intensity groups ($i=2$--6).

Table 1 shows the best-fitting shift factors and the log-normal average
values $E^{(i)}=\left\langle E_{{\rm p,max}}^{(i)}\right\rangle $ for our
selected intensity groups. The differences between them show the effect of
hardness/intensity correlation of gamma-ray bursts. Basically, the ratios
between the average double-peak energies for different brightness groups $%
E^{(1)}/E^{(i)}$ are very close to the best fitting shifting factors $Y^{(i)}
$. The difference of the average peak fluxes between the 100 brightest and
100 dimmest bursts is a factor of $\sim$43, while the the corresponding
log-normal average values of the double peak energy differ by a factor of $%
\sim$3 (Table 1).

\section{Equivalent time width, as a burst temporal parameter}

A robust temporal parameter for bursts is very difficult to define. Bursts
have a variety of light curves, and for many of them the light curves are
very complex. The evaluation of a {\it duration-type} parameter depends on
the sensitivity of the instrument, its energy range, and its time
resolution. The best known parameters $t_{50}$ and $t_{90}$ attribute
definite durations to any individual burst, but there are several
statistical biases that could affect the results when statistical studies
are performed using them.

Below we suggest another duration-type parameter that could be associated
with a group of bursts. It was defined using the Average Emissivity Curves (%
\cite{Mit4}). The ACE is known to be a quite robust signature for any large
group of bursts. This ACE profile represents the slow component of burst
variability, which can be interpreted as the {\it general envelope} of
individual light curves. By its nature, the ACE averages over the faster
variability, leaving only the signature of {\it slow clocks}, so in
computing the ACE of BATSE gamma-ray bursts we can use the DISCLA and CONT
data, which have 1024~ms and 2048~ms time resolution, respectively. The ACE
intermixes individual events with all their particular time profiles, and
represents them by one profile, which is a single asymmetric peak with
steeper rise front and flatter back slope. Therefore, for each intensity
group $i$ the ACE profile can be used to estimate the average duration of
rise fronts and back slopes for averaged events.

Although the BATSE data set is large, the variety of burst time profiles is
such that we must consider its effect on the variance among finite samples.
We have found by direct comparison of ACE profiles for different samples of
BATSE bursts that they were much more distinct than would be expected from
the errors of sample variance for individual samples. This means that a
random sample of bursts for individual groups does not ensure the
well-weighted contribution of events with all kinds of profiles. To study
the random choice statistics of bursts, a special Monte Carlo simulation was
performed using the total set of 603 bursts in the 3B Catalog with durations 
$t_{90}>2$~s (\cite{Mit5,Lit}). Indeed, the distributions of stretching
coefficients $Y$ due to statistics of random choice were found to be much
broader than expected from the sample variance predicted by normal
statistics. From the simulations, the 1$\sigma$ deviation around a
non-stretched value $Y\sim 1$ for a sample of N bursts was found to be 
\begin{equation}
\frac{\delta Y}Y = 0.13\cdot \sqrt{\frac{100}N } . 
\end{equation}
This value can be used as an estimate of the error in stretching factors
between ACEs for any two groups of $N$ bursts. The present size of our
selected intensity groups (Table 1) allows us to resolve stretching at $\sim
3\sigma$ significance between them provided the effect is larger than $\sim
1.41$ (\cite{Mit5,Lit}).

There is a simple analytic form that provides a very good fit to the ACE
profile $\Phi$(t) for time intervals of 20--50~s around the maximum: 
\begin{equation}
\Phi(t)=\left(\frac{t_0}{t_0+|t-t_{{\rm max}}|}\right)^a , 
\end{equation}
where the exponent $a$ has different values $a_{RF}$ and $a_{BS}$ for the
rise front and back slope, respectively. This law allows us to take into
account the energy dependence of ACEs (\cite{Mit3}) by interpolating in
energy between the parameters $a_{{\rm RF}}$, $a_{{\rm BS}}$ and $t_0$
measured for ACEs in three BATSE discriminator channels (25--50 keV, 50--100
keV and 100--300 keV). It also allows us to use a time-efficient procedure
to estimate the relative time-stretching factor between ACE profiles for any
two samples with different intensity (\cite{Mit5}).

The procedure to build the ACE includes the selection of the highest peak of
each burst $C_{max}$ {\it in count space} and the normalization of time
profiles by the $C_{max}$ value. Therefore, the ACE is sensitive to a bias
resulting from domination of positive fluctuations in the selected peaks.
Due to this bias the ACE profile is systematically lower in both wings;
i.e., the measured value is narrower than the true value. The bias is
stronger for dimmer bursts, where the influence of positive fluctuations is
larger. To take it into account, our reference group 1 (the brightest one)
was transformed into an artificial reference group by Monte Carlo
noisification, having the same event profiles but with dimmer peak fluxes.
When the reference group 1 was dimmed down to the level of the dimmest group
6, the noise-produced ACE was found to be different than the original ACE by
a factor of $\sim$0.8 (that is: narrower -- see \cite{Mit5}).

To estimate the average equivalent time width for the brightest reference
group 1, the observed ACE profiles in three energy ranges $j=1$ (25-50 keV), 
$j=2$ (50-100 keV) and $j=3$ (100-300 keV) were used as they are. The bias
due to positive fluctuations is assumed to have no influence on this group.
Using the ACE$^{(1,j)}$ profiles, we calculated average equivalent widths in
each of three energy channels $j$ as 
\begin{equation}
\tau _{{\rm RF,BS}}^{(1,j)}=\int dt\Phi _{{\rm FR,BS}}^{(1,j)} (t) 
\end{equation}
for the rise front (RF) and back slope (BS), respectively. The values of $%
\tau ^{(1,j)}$ were then interpolated over the broad energy range 25--300
keV, and the equivalent width was determined at the double-peak energy $%
E^{(1)}=293$~keV. The values of $\tau ^{(1,j)}(E^{(1)})$ for rise front and
back slope are the temporal parameters $\tau ^{(1)}_{{\rm RF,BS}}$ for the
reference group 1 (Table 1). Physically, $\tau$ represents the average
duration of emission either over the rise or over the decay at the spectral
range around the double-peak energy $E_{{\rm p,max}}$.

ACE$^{(i,j)}$ profiles for dimmer groups ($i=2$--6) can be used similarly to
estimate temporal parameters, provided that they are corrected for the
noise-produced narrowing of ACEs. For a given intensity group $i$ an
artificial reference group $i^{\prime }$ was created from the events of the
reference group 1 by Monte Carlo noisification, in which the reference
bursts are reduced in intensity to the fluxes of group $i$, and noise added
corresponding to the noise level of group $i$. The artificial reference
group $i^{\prime }$ therefore represents the original group 1, but takes
into account the noise-produced effects. Therefore, to evaluate the
noise-corrected average stretching between the testing group $i$ and the
reference group 1 for energy channel $j$, we measured the stretching factors 
$Y^{(i,j)}$ between the ACE$^{(i,j)}$ of the actual dim group $i$ and the ACE%
$^{(i^{\prime },j)}$ of the artificial reference group $i^{\prime }$. We
calculated $Y^{(i,j)}$ separately for rise fronts (RF) and back slopes (BS)
for intensity groups $i=2$--6 in three energy ranges $j=1$--3. Using these
factors , the parameters of equivalent width of ACE$^{(i,j)}$ could be
defined for the energy channels $j$, as 
\begin{equation}
\tau_{{\rm RF,BS}}^{(i,j)}=\tau_{{\rm RF,BS}}^{(1,j)}\cdot Y_{{\rm RF,BS}%
}^{(i,j)} . 
\end{equation}
These values are corrected for the noise-produced narrowing of ACEs because
they are determined from the stretching factors between the testing dim
groups and the corresponding artificially noisified reference groups. The
single-value temporal parameters $\tau ^{(i)}(E^{(i)})=\tau ^{(i)}$ were
interpolated between the values $\tau _{{\rm RF,BS}}^{(i,j)}$ for three
discriminator channels $j=1$--3 at the double-peak energies $E_{{\rm p,max}%
}^{(i)}$. The parameters $\tau ^{(i)}$ are presented in Table 1 for rise
fronts (RF) and back slopes (BS) of bursts. Their errors are estimated from
the choice statistics for stretching factors $Y_{i,j}$. The rise front
equivalent widths $\tau _{{\rm RF}}^{(i)}$ do not show a correlation with
burst intensity. On the other hand, the back slope equivalent widths $\tau _{%
{\rm BS}}^{(i)}$ are significantly increasing with decreasing intensity of
bursts.

\section{Average Cosmological Invariant Parameters}

The average double-peak energy $E_{{\rm p,max}}^{(i)}$ and the equivalent
time width $\tau ^{(i)}$ at the double-peak energy are very useful
parameters for testing the cosmological paradigm. Indeed, the
energy-dimension parameters $E_{{\rm p,max}}^{(i)}$ represent {\it spectral
signatures} that have the same physical sense for all bright, medium and dim
groups of bursts. The time-dimension parameters $\tau ^{(i)}$ represent {\it %
temporal signatures} that are also well-defined for all groups of bursts.

We have found that these parameters vary significantly among the different
intensity groups. However, the differences between them could be caused
either by the purely {\it geometrical} transformations of red shift and time
dilation in the expanding Universe, or by a {\it physical} variation among
the outbursting sources in the co-moving frames. These two parameters cannot
by themselves be used to perform a model-independent test of the
cosmological paradigm of gamma-ray bursts. One must either postulate some
intrinsic properties of emitters and then resolve the cosmological
transformations of observed gamma-ray bursts, or postulate the geometrical
effects of time dilation and energy red shift and then deconvolve properties
of observed bursts into the intrinsic properties of emitters.

As an alternative, we wish to find a special observational parameter for any
selected sample of gamma-ray bursts that does not depend on the geometrical
effects of the Universe extension, and which we call an {\it Average
Cosmological Invariant Parameter} (ACIP)(\cite{Mit1}). Let us assume that
some brightness group $i$ corresponds to emitters with red shifts around
some average value $z^{(i)}$ and the corresponding equivalent width and
double-peak energy equal $\tau_*^{(i)}$ and $E_*^{(i)}$ in the co-moving
frame, respectively. Then, since a time-dimensional average parameter is
increased by a factor $(1+z^{(i)})$, giving $\tau^{(i)} = \tau_*^{(i)}
(1+z^{(i)})$ in the observer's frame, and an energy-dimensional average
parameter is reduced by a factor $(1+z^{(i)})^{-1}$, giving $E^{(i)} =
E_*^{(i)} (1+z^{(i)})^{-1}$ in the observer's frame, the product $\Pi^{(i)}$
of time-dimensional and energy-dimensional average parameters for the group
is an invariant because the red shift factors cancel each other: 
\begin{equation}
\Pi^{(i)} = \tau^{(i)} \cdot E^{(i)} = \tau_*^{(i)} \cdot E_*^{(i)}. 
\end{equation}
Therefore, any difference between values of $\Pi$ for two different samples
of bursts has to be attributed to a real physical difference between their
emitters.

\section{Results from a comparison of ACIPs for different intensity groups}

The average durations of rise fronts and back slopes are known to have
different behaviors for bursts with the different intensity. We calculated $%
\Pi$ as defined in equation (6) for the same 6 brightness groups as in Table
1, separately for time signatures $\tau _{{\rm RF}}$ and $\tau _{{\rm BS}}$.
The results are presented in Table 2 and Figures 2 \& 3.

One can consider emitters of gamma-ray bursts as {\it standard candles} with
respect to the property described by the ACIP if the values are independent
of brightness. This model can be rejected for the rise fronts of bursts: the
assumption of a constant $\Pi_{{\rm RF}}$ in our analysis has a negligibly
small probability ($<0.001$). Indeed, $\Pi_{RF}$ decreases with decreasing
average fluxes $\left\langle F_{{\rm max}}\right\rangle $, as 
\begin{equation}
\label{ACIP}\Pi_{{\rm RF}}=1196\cdot \left[ \frac{\left\langle F_{{\rm max}%
}\right\rangle } {14.2}\right] ^{0.22\pm 0.03} ,
\end{equation}
where $\langle F_{{\rm max}}\rangle$ is in units of $\gamma$/cm$^2$-s and $%
\Pi_{{\rm RF}}$ is in units of keV-s. Therefore $\Pi_{{\rm RF}}$ cannot be
used as a {\it standard candle} (Figure 2). Of course, there may be a
concern that the dimmest group suffers from incompleteness, since it is
closest to the trigger threshold where slow-rising events can be missed (%
\cite{Kommers}). If we exclude the dimmest group from the fit in Figure 2,
the fitted power-law index in eq.~\ref{ACIP} does not change dramatically: $%
0.18 \pm 0.03$. In this case, the probability that the data are consistent
with a constant is still very small ($0.003$).

On the other hand, the values of $\Pi_{{\rm BS}}$ are consistent with a
constant value (Figure 3). Quantitatively, we find that 
\begin{equation}
\Pi_{{\rm BS}}=1350\cdot \left[ \frac{\left\langle F_{{\rm max}%
}\right\rangle } {14.2}\right] ^{0.03\pm 0.03} , 
\end{equation}
where the units are the same as in equation (7). During the back slopes, the
differences between average $E^{(i)}$ and average $\tau _{{\rm BS}}^{(i)}$
(Table 1) for different intensity groups $i=1$--6 effectively compensate
each other when they form such a product as $\Pi_{BS}$. Since the two are
physically different parameters of emitters, we conclude that their
brightness dependencies have predominantly a geometrical origin, i.e., they
are due to the geometrical transformations of time and energy in the
expanding Universe.

Comparing the values of $\langle E_{{\rm p}}\rangle$ and $\tau_{{\rm BS}}$
for the 100 brightest bursts (group 1, with peak fluxes $>3.8$ $\gamma$/cm$^2
$-s) with those for the 100 dimmest bursts (group 6, with peak fluxes $<0.43$
$\gamma$/cm$^2$-s), we find that the factor of cosmological transformation
between emitters of these groups, both for time dilation and red shift, is
about 3 (Table 1). For this factor the value of $z_{{\rm dim}}$ for emitters
of dimmest bursts is about 2, provided the brightest bursts correspond to $%
z_{{\rm br}}\ll 1$. Recently it has been suggested that $z_{{\rm br}}>1$,
based on measurements of the spectra of bright burst (\cite{Dez}), and as
the consequence of the idea that GRBs sources should follow the history of
star formation (\cite{Wij2,Che}). In this case the group of dimmest bursts
would have an average red-shift factor $z_{{\rm dim}}\sim (3\cdot (1+z_{{\rm %
br}}) - 1)$ as large as $\sim 5$. While the outbursting sources are
effectively standard candles along the back slopes, they are not standard
along the rise fronts of bursts. If the bursts have a cosmological origin,
observations of dim and bright bursts correspond in local time to the
younger and older Universe, respectively. The variation of $\Pi_{{\rm RF}}$
with intensity (Table 2) is associated with a difference of average duration 
$\tau _{{\rm RF}}$ in the co-moving frames of reference, because the values
of $E_{{\rm p,max}}$ are the same for both $\Pi_{{\rm BS}}$ and $\Pi_{{\rm RF%
}}$. In the co-moving frames the bursts in the recent Universe have an
average rise time $\sim$3 times shorter than the average rise time of bursts
in the early Universe.

The difference in rise time between emitters of bright and dim bursts could
be the result of differences in the interaction of the outbursting source
with the surrounding medium. Emitters of dimmer bursts could have interacted
in a medium with higher density, or with a harder background emission, or
with a stronger average magnetic field, or with a difference of some other
global parameter of the Universe. On the other hand, during the back slope
there is no difference between bursts from close and distant cosmological
distances. The tails of bursts are thought to represent a self-determined
internal process, which has some internal time scale associated either with
some inertia, or with a time constant of some decay, or with some other
process, and which does not depend on the external condition of the
surrounding medium. A future cosmological model has to take into account
these differences between processes of emission during the rising phases of
bursts and their decays, which is apparently intrinsic to the co-moving
reference frames.

\clearpage

\figcaption{The $E_{\rm p,max}$ distributions for different brightness groups.
The distribution for the reference group is showed by thick line. The
distribution for the dimmest group (6) is showed by thin line.\label{Fig1}}

\figcaption{The dependence of $ACIP_{\rm RF}$ on $\left\langle F_{\rm
max}\right\rangle$. The best linear fit (solid curve) has a non-nzero
slope.\label{Fig2}}

\figcaption{The dependence of $ACIP_{\rm BS}$ on $\left\langle F_{\rm
max}\right\rangle$. The best linear fit (solid curve) has a slope consistent
with zero.\label{Fig3}}

\begin{deluxetable}{lcccccc}
\tablecaption{}
\label{table1}
\tablehead{
\colhead{Intensity} &
\colhead{Peak flux} &
\colhead{Average peak flux} &
\colhead{$\left\langle E_{\rm p}\right\rangle$} &
\colhead{$Y$} &
\colhead{$\tau_{\rm RF}$} &
\colhead{$\tau_{\rm BS}$}\\
group &  $(\gamma$/cm$^{2}$s) & $(\gamma$/cm$^{2}$s) & (keV) & & (s) &(s)}
\startdata
1 &  $>3$.8 &14.2$\pm 1.4$ & 293$\pm 23$ & $1.0$& 4.1$\pm 0.5$ & 4.6$\pm
0.6$ \nl
2 &  1.6-3.8 & 2.4$\pm 0$.3 & 235$\pm 21$ & $1.22\pm 0.12$ & 4.7$\pm 0$.6 &
7.3$\pm 0$.9 \nl
3 &  0.95-1.6 & 1.2$\pm 0$.1 & 160$\pm 12$ & $1.82\pm 0.35$ & 4.9$\pm 0$.6
& 8.8$\pm 1$.1 \nl
4 &  0.62-0.95 & 0.77$\pm 0$.08 & 134$\pm 14$ & $2.07\pm 0.20$ & 4.6$\pm
0$.6 & 8.0$\pm 1$.0 \nl
5 &  0.43-0.62 & 0.51$\pm 0$.05 & 116$\pm 12$ & $2.58\pm 0.23$ & 5.3$\pm
0$.7 & 10.2$\pm 1$.3 \nl
6 &  $<0$.43 &  0.33$\pm 0$.03 & 97$\pm 10$ & $2.92\pm 0.10$ & 4.5$\pm 0$.6
& 12.4$\pm 1$.6 \nl
\enddata
\end{deluxetable}

\begin{deluxetable}{lcc}
\tablecaption{}
\label{table2}
\tablehead{
\colhead{Intensity} &
\colhead{$\Pi_{\rm RF}$} &
\colhead{$\Pi_{\rm BS}$} \\
group & (keV-s) & (keV-s)}
\startdata
1 & $1196\pm 174$  & $1350\pm 205$ \nl
2 & $1107\pm 172$  & $1716\pm 271$ \nl
3 & $784\pm 113$   & $1402\pm 211$ \nl
4 & $616\pm 103$   & $1072\pm 179$ \nl
5 & $613\pm 103$   & $1183\pm 194$ \nl
6 & $438\pm 74$    & $1203\pm 199$ \nl
\enddata
\end{deluxetable}

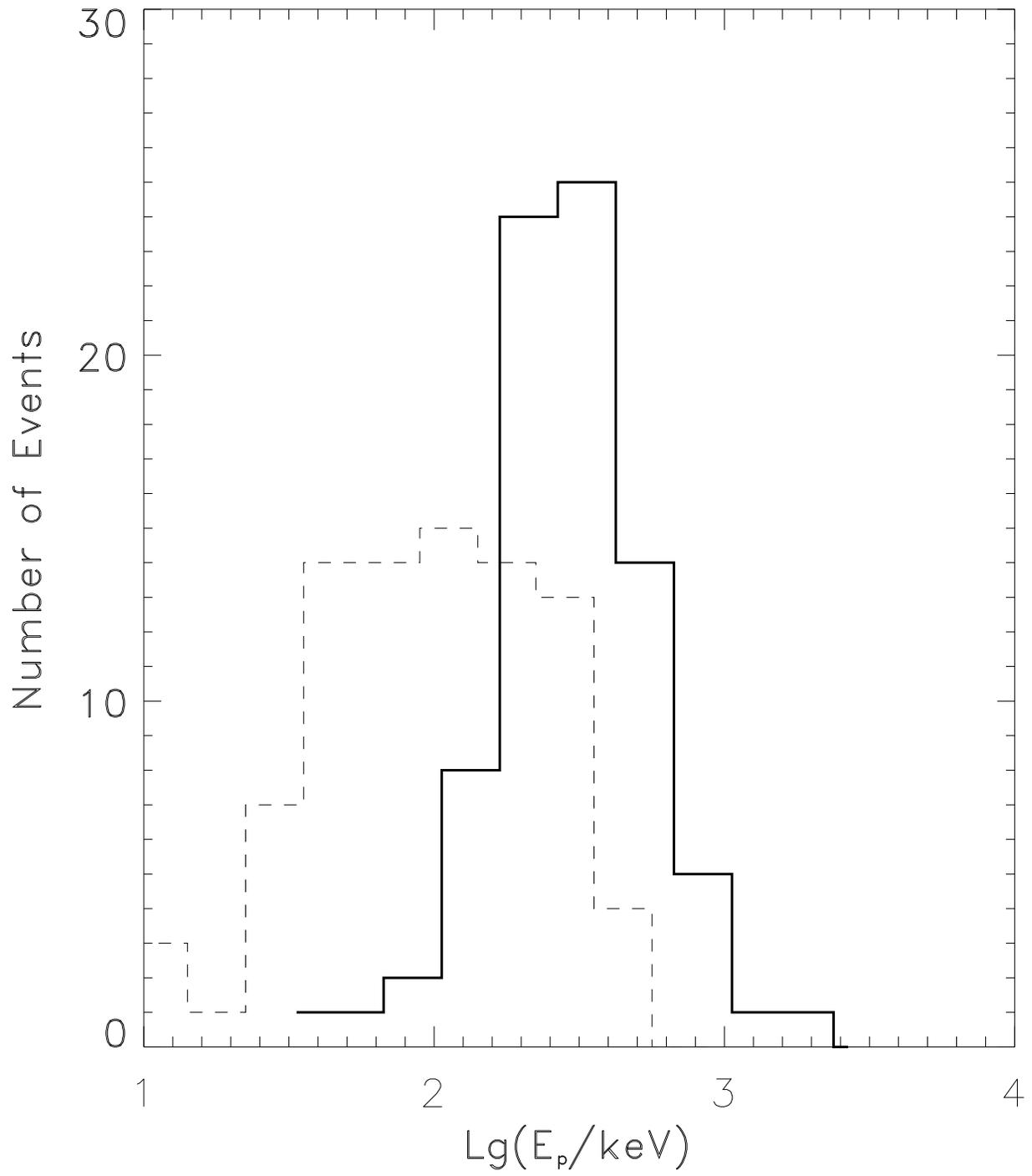
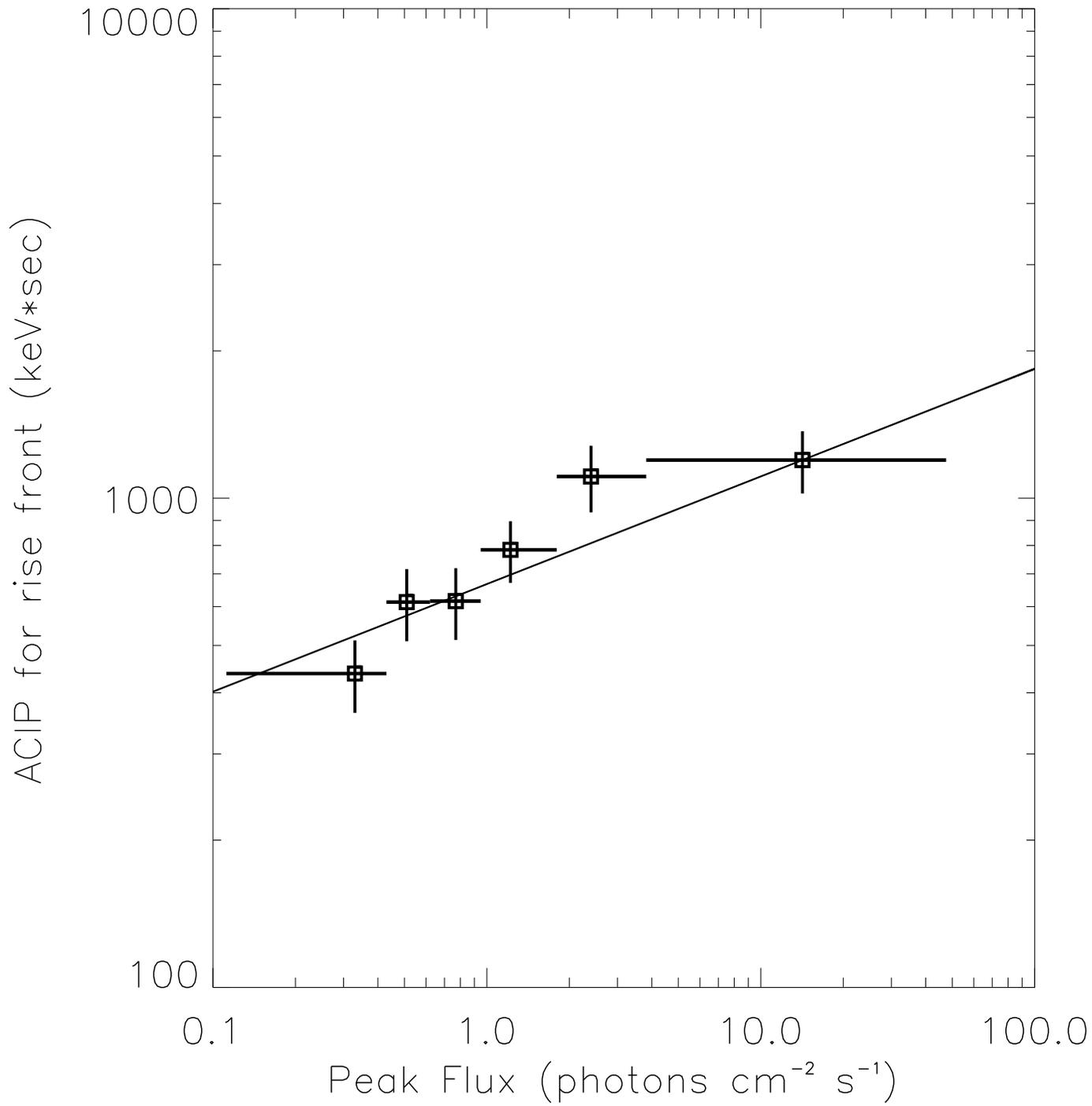
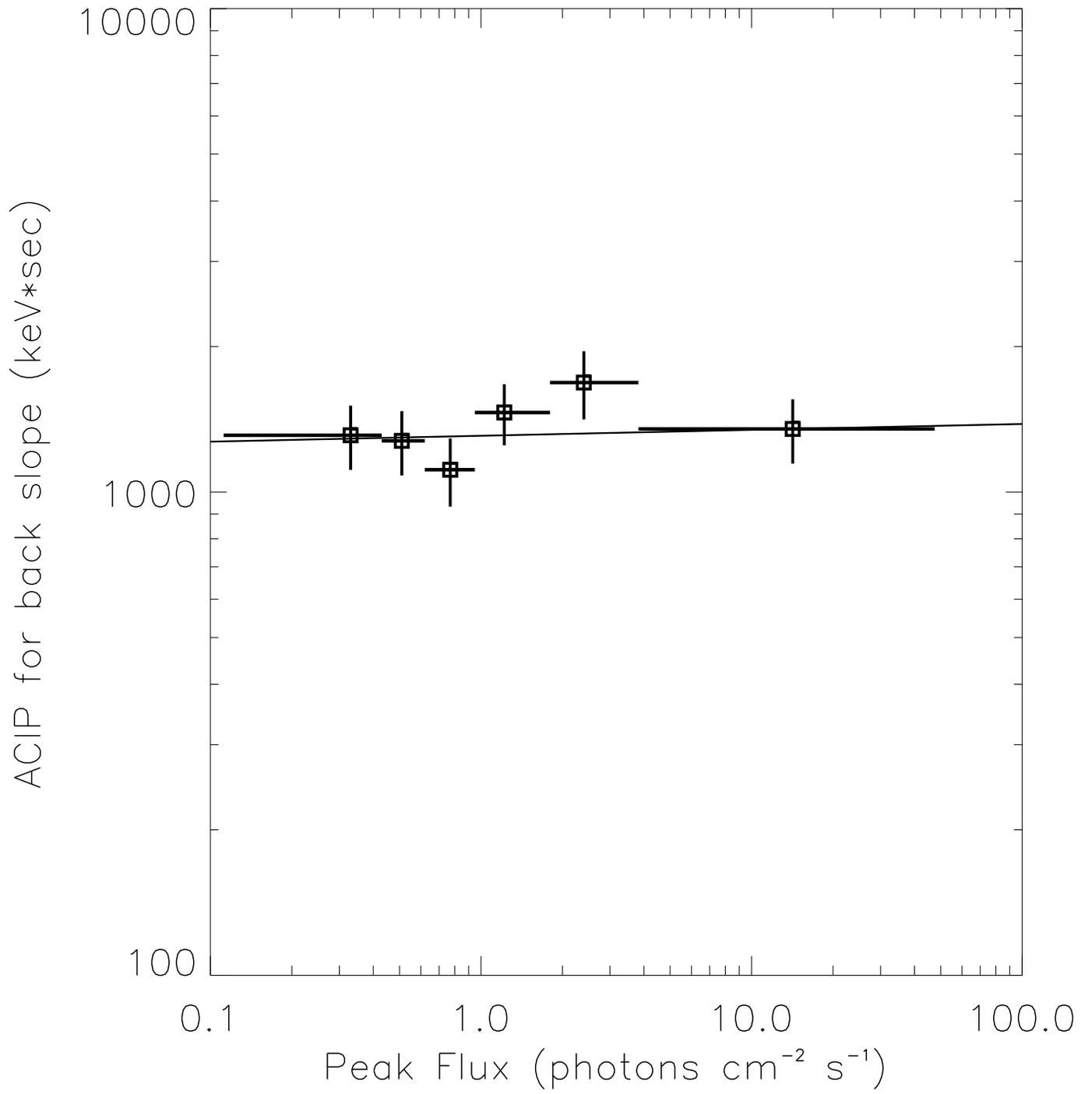
\end{document}